\documentclass[prb,twocolumn,floatfix,groupedaddress]{revtex4}
\usepackage[T1]{fontenc}
\usepackage[english]{babel}
\usepackage{amsmath}
\usepackage{graphicx}
\newcommand{\nd}{\noindent}
\newcommand{\be}{\begin{equation}}
\newcommand{\ee}{\end{equation}}
\newcommand{\ben}{\begin{eqnarray}}
\newcommand{\een}{\end{eqnarray}}
\begin{document}

\title{Scale-invariance underlying the logistic equation and its social applications}

\author{A. Hernando$^1$, A. Plastino$^{2,\,3}$}
\address{$^1$ Laboratoire Collisions, Agr\'egats,
R\'eactivit\'e, IRSAMC, Universit\'e Paul Sabatier
118 Route de Narbonne 31062 - Toulouse CEDEX 09, France\\
$^2$National University La Plata, IFLP-CCT-CONICET, C.C. 727, 1900
La Plata, Argentina   \\ $^{3}$ Universitat de les Illes Balears
and IFISC-CSIC, 07122 Palma de Mallorca, Spain}

\begin{abstract}
On the basis of dynamical principles we derive the Logistic
Equation (LE), widely employed (among multiple applications) in
the simulation of population growth, and demonstrate that
scale-invariance and a mean-value constraint are sufficient and
necessary conditions for obtaining it. We also generalize the LE
to multi-component systems and show that the above dynamical
mechanisms underlie large number of scale-free processes. Examples
are  presented regarding city-populations, diffusion in complex
networks, and popularity of technological products, all of them
obeying the multi-component logistic equation in an either
stochastic or deterministic way. So as to assess  the
predictability-power of our present formalism, we
 advance a prediction, regarding the next 60 months, for
the number of users of the three main web browsers (Explorer,
Firefox and Chrome) popularly referred as ``Browser Wars''.
\end{abstract}
\pacs{89.70.Cf, 05.90.+m, 89.75.Da, 89.75.-k}
\maketitle

\section{Introduction}

\nd It is well-known that the logistic equation (LE) (sometimes
called the Verhulst model or logistic growth curve) is a
phenomenological model of population growth first published by
Pierre Verhulst in the 1840's. The model is continuous in time,
but a modification of the pertinent equation to a discrete
quadratic recurrence equation known as the logistic map is also
widely used. The continuous version of the logistic model for the
evolution of the population $x(t)$ is described by the
differential equation \be\label{le} \dot{x}(t)= k
x(t)\left(1-\frac{x(t)}{N}\right), \ee where $k$  is the
Malthusian parameter (rate of maximum population growth) and $N$
is the so-called carrying capacity (i.e., the maximum sustainable
population).  The LE has as a solution \be\label{sle} x(t)=
\frac{N}{1+(N/x(0)-1)e^{-kt}}, \ee i.e., the sigmoid function. The
discrete version of the LE is the celebrated logistic map. A
typical application of the logistic equation refers to  a
1838-model of population growth, originally due to Verhulst, in
which the rate of reproduction is proportional to both the
existing population and the amount of available resources, all
else being equal. The Verhulst equation was published after he had
read Thomas Malthus' {\it An Essay on the Principle of
Population}. Verhulst derived his logistic equation to describe
the self-limiting growth of a biological population. Today, proper
referencing to the logistic equation's variegated applications to
multiple fields of endeavor would require pages and pages of
citations. Of this immense body we just mention, as a tiny sample,
\cite{1,2,3,4,5,6,7}.

\nd Some ad-hoc LE-deductions have been previously published in a
case-by-case basis. One such demonstration is that provided by A.
D. Zimm for companies or firms \cite{zimm}, that grow according to
their commercial successfulness with a classic linear Marshallian
price-volume relationship. Other analytical derivations are also
found with some ad-hoc assumptions\cite{6,toston}. {\it Out
present goal is to describe an universal and generic dynamical
mechanism} that leads in natural fashion to the logistic equation.
The above cited derivations do not have a dynamical character, as
ours does. Our procedure is based on \begin{itemize} \item
\emph{scale symmetry} and \item a \emph{mean-value constraint}.
\end{itemize} We will   show that these are necessary an sufficient conditions
for a LE-derivation. The two items above are empirically known to
be {\it related to the LE} \cite{zimm,6,toston} but they are
used here for the first time as its {\it pure mathematical basis}.

\section{Derivation from dynamical principles}

\nd Consider an $n-$components system, each of them characterized
by a population $x_i$. Let us further assume that a
multiplicative, time-evolution of population takes place via free
proportional growth, i.e.,
$$\dot{x}_i(t) = k_i(t) x_i(t),$$
where $k_i$ is the growth-ratio per unit-time for the $i$-th
component. Scale-symmetry is here apparent so that it proves
convenient to transform coordinates to $u_i=\log(x_i)$ as in Refs.
\onlinecite{prl,epjb}. Thereby one is led to the linear equation
$$\dot{u}_i(t) = k_i(t),$$
where the scale-invariance of $x$ is now a translational
invariance in $u$. We assume that the total population is finite,
namely
$$\sum_{i=1}^{n} x_i(t) = \sum_{i=1}^{n} e^{u_i(t)} = N, $$ and
that also  $n$ remains constant, so that  $\langle x\rangle=N/n$.
For each arbitrarily small time-interval $\Delta t$, the
$u-$population  grows freely via
$$u'_i(t+\Delta t) = u(t) + \Delta t k_i(t),$$
but conservation of the mean value $\langle e^u\rangle$ demands a
global ``self-correction'' process that should  respect the
original symmetry of the system (translational for $u$).
Accordingly,
$$u_i(t+\Delta t) = u_i'(t+\Delta t) + A,$$
where $A$ is a value that guarantees fulfillment of
$\sum_{i=1}^{n} e^{u_i(t+\Delta t)} = N$. This is achieved if
$$A = - \ln\left[\frac{1}{N}\sum_{j=1}^{n}\, e^{u_j'(t+\Delta t)}\right]= - \frac{\Delta t}{N}\sum_{j=1}^{n} k_j(t)e^{u_j(t)},$$
where a Taylor-expansion to first order is justified since $\Delta
t$ is arbitrarily small. One is then led to
$$u_i(t+\Delta t) = u_i(t)+\Delta t\left[k_i(t)-\frac{1}{N}\sum_{j=1}^{n} k_j(t)e^{u_j(t)}\right].$$
In the  continuum-limit  one finds
$$\dot{u}_i(t) = k_i(t)-\frac{1}{N}\sum_{j=1}^{n} k_j(t)e^{u_j(t)},$$
that written in $x-$parlance leads to what we call the
\emph{multi-component logistic equation} (MCLE) \be\label{mcle}
\dot{x}_i(t) = x_i(t)\left(k_i(t)-\frac{1}{N}\sum_{j=1}^{n}
k_j(t)x_j(t)\right). \ee A matrix version of this equation is
presented in the Appendix, together with with some applicability
perspectives. It is easy to check that the MCLE retains the
original scale-symmetry of $x$, and also exhibits translational
symmetry in $k$. The latter allows for some arbitrariness in the
definition of the $k_i$ rates in the fashion $k'_i=k_i-k_0$. The
same results obtain for primed or unprimed $k$'s. If the $k_i$ are
constant, or  exhibit just a slow dependence on $t$
(quasi-statics), the solution to the MCLE is \be\label{smcle}
x_i(t) = \frac{N x_i(0)e^{k_it}}{\sum_{j=1}^{n}x_j(0)e^{k_jt}},
\ee where $x_i(0)$ are the initial conditions of the
evolutive-process. The logistic equation is directly derived from
the MCLE Eq. (\ref{mcle}) in a straightforward fashion. We recover
Eq. (\ref{le}) by i) considering a bi-component system ($n=2$),
ii) defining $x(t)\equiv x_1(t)$, $k\equiv k_1-k_2$, and iii)
taking into account that $x_2(t)=N-x_1(t)$. Similarly, the sigmoid
function Eq. (\ref{sle}) is recovered from Eq. (\ref{smcle}) with
the same assumptions. The second component acts here as a
population-reservoir and the first component becomes the only
evolutive degree of freedom.

\section{Possible physical regimes}

\nd According to the nature of the growth-ratios $k_i$, variegated
kinds  of processes can be described by the MCLE.
\begin{enumerate} \item
 Constant
values or deterministic time-dependencies lead to mechanical
systems exhibiting deterministic evolution while \item  adding
noise to the pertinent  mean values gives rise to stochastic
systems, with interesting behaviors and
applications.\end{enumerate} Without aiming to be exhaustive, we
consider here three different tableaus for MCLE-applicability,
according to the amount of `noise' in the system: totally
stochastic (or thermodynamic regime), an intermediate level of
randomness (involving diffusive processes), and totally
deterministic dynamics.

\subsection{Thermodynamic regime}

\nd Consider  a multi-component case with dozens or hundreds of
elements, and a very high level of noise. Assume that each $k_i$
describes the derivative of a Wiener process. We write
$k_i(t)=\overline{k}_i + \sigma_i \xi(t)$, where $\overline{k}_i$
is the time-average of $k_i$, $\sigma_i$ the standard deviation
measured in a certain interval $\Delta t$, and $\xi(t)$ an
independent normal-distributed random number. Defining
$\langle\sigma^2\rangle=\sum_{i=1}^n\sigma_i^2/n$, one asserts
that we {\it thermalize} the system if i)
$|\sigma_i-\sigma_j|^2/\langle\sigma^2\rangle\ll1$ and ii)
$|\overline{k}_i-\overline{k}_j|^2/(\langle\sigma^2\rangle\Delta
t)\ll1$, $\forall i,j$, i.e., if all elements exhibit similar
deviations and the differences between mean values are much
smaller than the noise. If $n$ is large enough (as stated above,
in the hundreds), dynamical equilibrium is encountered after some
finite time, meaning that the system is well-described by the
MaxEnt approach. The MaxEnt solution for scale-free systems
describes an equilibrium density $p_X(x)$ that follows the general
form \cite{prl}
$$
p_X(x)dx=\exp\left[-\sum_a f_a(x)\right]\frac{dx}{x},
$$
where $f_a(x)$ is the $a$-th constraint of the system. For a
constraint in the normalization ($n$ is invariant) we have
$f_{\langle 1\rangle}(x)=\mu$, and for one in the mean value we
write $f_{\langle x\rangle}(x)=\lambda x$, where $\mu$ and
$\lambda$ are constants, univocally determined by the fulfillment
of each constraint. The density-distribution obeys in this case
the relation \cite{epjb}
$$
p_X(x)dx=\frac{1}{\Gamma(0,\lambda x_0)}\frac{e^{-\lambda x}}{x}dx;~~x_0\leq x<\infty,
$$
where $\Gamma$ is the incomplete gamma function and $x_0$ is the smallest allowed
population for the elements (that can be 1). The rank-distribution is then written as
\be\label{merd}
x = \frac{1}{\lambda}\Gamma^{-1}\left[0,\Gamma(0,\lambda x_0)r/n\right],
\ee
where $r$ is the (continuous) rank from 0 to $n$, and
$\Gamma^{-1}(z)$ denotes the inverse function of $\Gamma:$ $\Gamma(\Gamma^{-1}(z))=z$.
The value of $\lambda$ is obtained from the mean value
\be\label{lambda}
\frac{e^{-\lambda x_0}}{\lambda\Gamma(0,\lambda x_0)}=\frac{N}{n}.
\ee

In order to test the above relationships we have carried out a
calculation that simulates human population dynamics. We also
compare the result with empirical data regarding city and
place-populations, using as example data from  Ohio State (United
States) \cite{ohio}. We consider $n=1000$ random walkers that
mimic the population of cities, and set the total population to
$N=6000000$. We fix the minimum allowed size at the Dunbar's
number \cite{epjb} $x_0=150$ people (empirically known to be the
usual size of small human communities, and related to the maximum
of social relationships/links that a human can comfortably
handle). We make the walkers to stochastically obey the MCLE Eq.
(\ref{mcle}) so as to simulate migration patters between  cities,
with a constant total population. We evolve the walkers in small
intervals $\Delta t=0.03$ and generate gaussian-distributed random
numbers at each interval for each $k_i$, using $\sigma_i=1$ for
all $i$. Due to the translational symmetry in $k$, we can set
$\overline{k}_i=0$ for all $i$. To respect the minimum size, a
walker' `move' is not accepted if it leads to a value lower than
the Dunbar's one. We start all walkers at $x_i=N/n$. After some
iterations we get the equilibrium distribution of Fig. 1, which
perfectly fits the MaxEnt prediction ($\lambda=0.00533$
humans$^{-1}$). The available empirical data covers years 2010,
2000 and 1990 with $n=1204$, $1015$, and $928$ cities and places,
respectively. We discard places with populations of under 150
people (67, 48 and 39 centers respectively). We also discard very
large cities (their potential economic correlation with the rest
of the country compromises the hypothesis of isolated systems).
Excluding  the four largest cities, the total population is
$N=6318170$, $6019960$, and $5477830$, respectively. We have
checked the proportional growth condition comparing $\log(x_i)$
vs. $|\dot{x}_i/x_i|^2$ (or equivalently, $u_i$ vs.
$|\dot{u}_i|^2$). No correlation between these two observables is
expected for scale-invariance. We have found a correlation
coefficient of $0.0018$, as shown in Fig. 1, thus confirming the
proportional growth hypothesis (the same correlation coefficient
in the precedent simulation is $0.0027$). According to Eq. (\ref{lambda}),
the predicted values of $\lambda$ are $0.00585$, $0.00507$, and
$0.00513$, respectively. A direct fit of the data to the form Eq.
(\ref{merd}) yields $\lambda=0.00636(2)$, $0.00502(3)$, and
$0.00522(3)$, respectively, close enough to the former values so
as to confirm the MaxEnt prediction.

\begin{figure}[ht]
\begin{center}
\includegraphics[width=\linewidth,trim =10pt 0pt 30pt 90pt,clip=true]{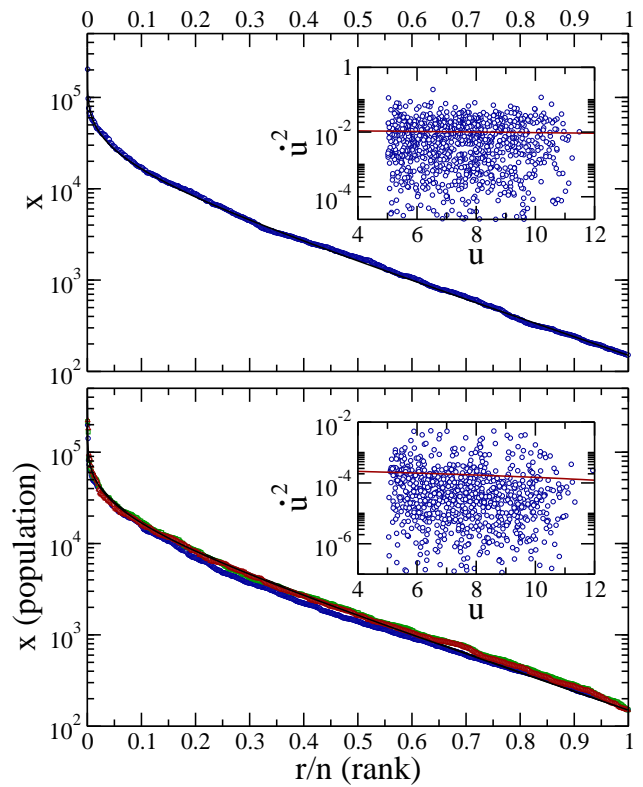}
\caption[]{Top panel: behavior of random walkers obeying the
stochastic MCLE (blue dots) compared with the associated MaxEnt prediction (black line).
Inset: walkers' squared-relative-growth $|\dot{u}_i|^2$ vs. logarithmic
size $u$. There is no correlation in view of the linear regression (red line).
Bottom panel: Rank-distributions for Ohio (red: year 2010; green:
year 2000; blue: year 1990) and rank-distribution of random
walkers obeying the stochastic MCLE (yellow) compared with the
corresponding MaxEnt predictions (black lines). Inset: same as upper panel
with the empirical data. In view of the regression line, no correlation is detected.}
\end{center}
\end{figure}

\subsection{Intermediate regime}

\nd  Let us pass now to consideration of an intermediate
stochastic regime in  the bi-component case. One party acts as a
population-reservoir while the other obeys Eq. (\ref{le}).
Defining the new variable \be\label{y} y(t)=-\log(N/x(t)-1), \ee
the LE linearizes itself for $\dot{y}(t)=k$. The rate $k$ is here
again the derivative of a Wiener process
$k(t)=\overline{k}+\sigma\xi$, but we now include a drift obeying
 $|\overline{k}|^2>\sigma^2\Delta t$. Working with  an ensemble of
independent walkers following this equation is equivalent to
handling Brownian motion in $y$-space. Consequently, the usual
diffusion equation for the density of walkers for $p_Y(y,t)$
ensues  \cite{BM}
$$
\partial_t p_Y(y,t) = -\overline{k}\partial_y p_Y(y,t) + D\partial_y^2 p_Y(y,t),
$$
where $D$ is the diffusion-coefficient, related to $\sigma_k$ via
$\sigma_k=\sqrt{2D/\Delta t}$, and $\Delta t$ stands for  the
time-interval used in the random-walk numerical simulation. The
diffusion-equation's kernel is a Gaussian
$$
p_Y(y,t)dy = \frac{1}{\sqrt{4\pi Dt}}e^{-\frac{(y-y_0-\overline{k}t)^2}{4Dt}}dy,
$$
with $y_0$ a reference-value. If at $t=0$ all walkers
are located in $x$-space at $x_0=N/(1+e^{-y_0})$, they will later evolve via
\be\label{pxy}
p_X(x,t)dx = \frac{1}{\sqrt{4\pi Dt}x(1-x/N)}e^{-\frac{(\log(N/x-1)-y_0+\overline{k}t)^2}{4Dt}}dx,
\ee
since $p_X(x,t)dx=p_Y(y,t)dy$.

\nd  We have verified this prediction with a diffusion process
taking place inside a scale-free ideal network (SFIN)
\cite{ournets}, a random network with a degree-distribution
following the scale-free ideal gas one $p(c)\sim c^{-1}$, where
$c\leq c_M$. We have generated a SFIN of $N=20000$ nodes with a
maximum degree of $c_M=100$-connections and carried out a
multitude of cluster-growth processes \cite{ournets,clusters}.
Diffusion in networks generally starts i) by using a randomly
chosen node as a seed, ii) with its first neighbors being added to
the cluster in the first iteration, iii) and   the neighbors of
those ``first" neighbors, afterwards (and so on). The process ends
when all nodes of the network belong to the cluster. The size of
the cluster at each iteration depends on the particular node
selected as seed, via its position inside the network. We depict
in Fig. 2 the result of a large number of these processes,
indicating the size of the cluster at each iteration. All of them
start with $x=1$ and end up with $x=N$, but  processes exhibit
deviations at intermediate steps. The associated  median clearly
follows a logistic evolution, as shown in Fig. 2. Changing to the
variable $y$ of Eq. (\ref{y}) we find a straight line with slope
$\overline{k}=3.09$. The deviations can be described by $y$-random
walkers with $\sigma=0.7$ ($\Delta t=1$). The statistics of the
processes can be nicely described with Eq. (\ref{pxy}) via
$x_0=1$, $D=0.245$, and with the above value of $\overline{k}$, as
illustrated by  the comparison depicted in Fig. 2.

\begin{figure}[ht]
\begin{center}
\includegraphics[width=\linewidth]{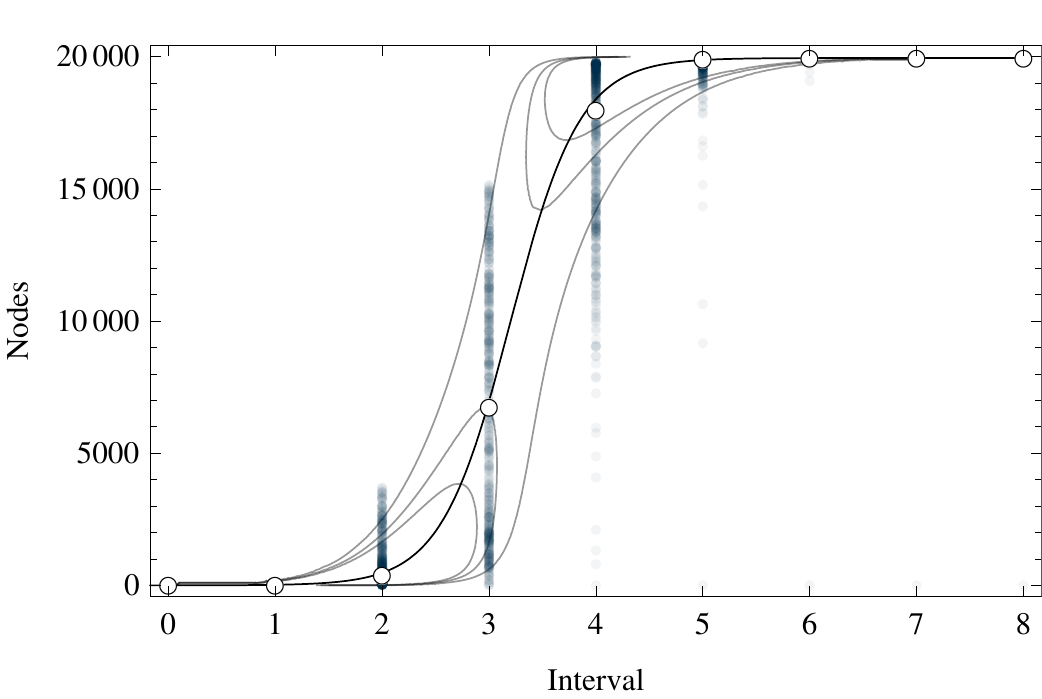}
\caption[]{Diffusion inside a scale-free ideal network (dots)
compared with the analytical one provided by Eq. (\ref{pxy}),
derived from the logistic equation.
 }
\end{center}
\end{figure}

\subsection{Deterministic  regime}

\nd  We study  now the deterministic evolution of a
multi-component system with few elements and very low level of (or
without) noise. Assume that the growth-ratios $k_i$ are now
constants or represent a quasi-static evolution in time. Assume
further  that we have data about the temporal population-evolution
but do not known the explicit values (or the tendency) of the
$k_i$ rates. These can be easily obtained from the solution of
MCLE Eq. (\ref{mcle}) taking advantage of its property of
translational symmetry in $k$. By arbitrarily setting $k_1=0$, all
the remaining values are obtained from the population data thanks
to the functions \be\label{ki} h_i(t) =
\log\left[\frac{x_i(0)}{x_i(t)}\frac{x_1(t)}{x_1(0)}\right]. \ee
If the growth-ratios are constants, $h_i(t)=k_i t$, the entire
evolution-path can be predicted (for any arbitrary time). If our
functions $h_i(t)$ exhibit small time-fluctuations we can
parameterize them, via a fitting procedure, to any given
analytical form. The desired solution is obtained by substituting
the arguments in the exponentials of Eq. (\ref{mcle}) by these
functions $k_i t\rightarrow h_i(t)$.  We have tested this last
statement using data regarding web-browsers' statistics so as to
study the past and future of the (popularly called) \emph{Browser
Wars} \cite{wikiBW}. We considered the $n=3$ system composed of
Microsoft Explorer (E), Mozilla Firefox (F), and Google Chrome
(C). Our analysis of the popularity of each uses data from
\emph{w3schools} \cite{w3} and \emph{statcounter} \cite{sc}
(depicted in Fig. 3).

\nd We take $N=100$\% and choose the origin $t=0$ at March 2012
(as this communication was being written). Setting $k_E=0$ we have
applied Eq. (\ref{ki}) to the data, finding a small dependence on
time in both $k_F$ and $k_C$. We show in Fig. 3 that the functions
$h_i(t)$ can be nicely fitted to a simple exponential form
$h(t)=ae^{-bt}t+c$ (that can be regarded as a pure exponential
time-dependence of $k$  plus a correction on the initial value
$x(0)$ via $e^c$). Note that the small fluctuations of the data
become more apparent as we approach the reference point at $t=0$.
However,  our accuracy remains sufficiently high  for our
purposes. We  obtain
\begin{equation*}
\begin{array}{rl}
h_F(t)=&0.0074(5) \exp[-0.021(1)t]t-0.008(14)~\mathrm{and}\\
h_C(t)=&0.0579(24)\exp[-0.0097(10)t]t+0.104(26),
\end{array}
\end{equation*}
for \emph{w3schools}, and
\begin{equation*}
\begin{array}{rl}
h_F(t)=&0.0022(5) \exp[-0.043(6)t]t+0.026(13)~\mathrm{and}\\
h_C(t)=&0.055(2)  \exp[-0.015(1)t]t+0.015(27),
\end{array}
\end{equation*}
for \emph{statcounter}, that are compared in the top panels of
Fig. 3 to empirical data. Our predictions for the popularity
evolution are evaluated using Eq. (\ref{smcle}) and substituting
$k_Et$, $k_Ft$, and $k_Ct$ by the above described extrapolations
of $h_E$, $h_F$, and $h_C$. A proper correction is finally added
in the later case to improve the fitting by using $N'=1.03N$. We
depict in Fig 3 our monthly prediction for the next 5 years
regarding browsers' usage.
 In the two reported instances, Google Chrome grows till coming ahead in the competition, saturating effects being noticeable
at 80\% and 60\%, respectively. Thus, according to our prediction,
Google Chrome will win the competition but it will not acquire
such a  dominant position as the MS Explorer attained in the past.

\begin{figure}[ht]
\begin{center}
\includegraphics[width=\linewidth,trim =10pt 0pt 30pt 340pt,clip=true]{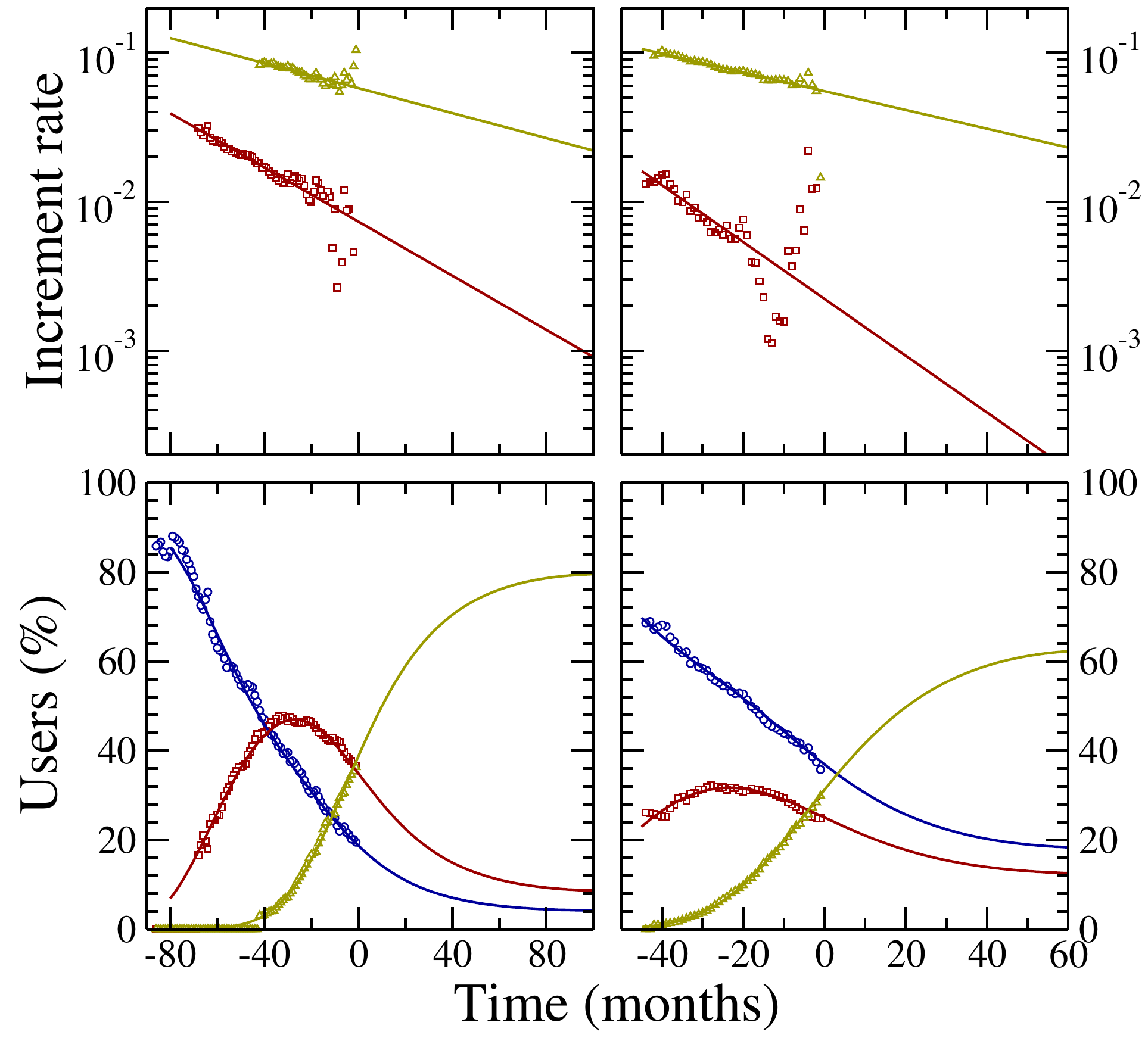}
\caption[]{
Left panels: data from \emph{w3schools}. Right panels:
data from \emph{statcounter}.
Top panels: increment rate of M Firefox (red squares) and G Chrome
(yellow triangles) defined as $(h(t)-c)/t$ relative to
MS Explorer (see text), compared with the analytical fit (solid lines).
Bottom panels: relative users of MS Explorer (blue circles),
M Firefox (red squares) and G Chrome (yellow triangles),
compared with our prediction (lines).
 }
\end{center}
\end{figure}

\section{Conclusions}

\nd We have been able here to demonstrated that
scale-invariance and a mean-value constraint are sufficient and
necessary conditions for obtaining the LE from first dynamical 
principles. Then, the LE was generalized to multi-component systems. This
allowed us to show that these dynamical mechanisms underlie
interesting scale-free processes, which was illustrated with
reference to  city-populations phenomena, diffusion in complex
networks, and popularity of Net Browsers.

\appendix
\section{Generalization of the multi-component logistic equation to a matrix equation}

\nd We generalize here the formalism discussed above. If we define
a new set of variable $\chi_i=\sqrt{x_i/N}$, the
total-population's constraint  can be recast in the fashion
$\sum_{i=1}^n\chi_i^2=1$. This condition becomes formally
equivalent to the conservation of the modulus of a vector
$\mathbf{\chi}=\{\chi_i\}_{i=1}^n$ in a $n$-dimensional space. We
can also generalize the definition of the growth-ratios $k_i$,
promoting them to a matrix  $K_{ij}=k_i\delta_{ij}$, and write the
MCLE as a matrix equation. Using bra-ket notation
$\mathbf{\chi}=|\chi\rangle$ one has
$$
\partial_t|\chi\rangle=\frac{1}{2}\left\{K-\frac{\langle\chi|K|\chi\rangle}{\langle\chi|\chi\rangle}\right\}|\chi\rangle.
$$
This equation is formally identical to that used in quantum
physics to find the ground state wave-function of a Hamiltonian
(here, $K$). We are speaking of the Imaginary Time Method (ITM)
widely used in the literature \cite{ITM}. The mean-value term is
also used to guarantee the conservation of the normalization of
$|\chi\rangle$ during the process. All our examples can be
regarded as particular applications of this formalism, calling
attention to  the ``functional" definition of our effective
`Hamiltonians' $K$. In our examples $K$ has a diagonal form, which
only indicates that we were working in the eigenbasis of the
dynamics defined by $K$. A generalized definition can include
off-diagonal terms as well, indicative of some kind of interaction
between populations. Density functional
theories (DFT) also use the above equation for many-body quantum
systems \cite{DFT}. A phenomenological Hamiltonian is defined by
means of a parametric functional form, than can also be a
functional of the own state-vector $\chi$. The associated
parameters  are chosen so as to reproduce well-known empirical
facts regarding  the system of interest. We expect that the bridge
we have here built up  between the MCLE and the DFT can open  new
vistas with respect to the possibility of  studying scale-free
systems. Such approach would take advantage of the huge experience
accumulated regarding DFT methods.


\begin{thebibliography}{}
\bibitem{1} S. Jannedy, R. Bod, J. Hay, {\it Probabilistic Linguistics} (MIT Press, Cambridge, Massachusetts, 2003).

\bibitem{2} N. A. Gershenfeld, {\it The Nature of Mathematical Modeling} (Cambridge University Press,  Cambridge, UK, 1999).

\bibitem{3} S. E. Kingsland, {\it Modeling nature: episodes in the history of population ecology} (University
of Chicago Press, Chicago, 1995).

\bibitem{4}  E. W. Weisstein, {\it Logistic Equation}, from Wolfram Research Mathworld, Repository hosted at UIUC.

\bibitem{5} M. Fuentes, H. Larrondo, M. T. Martin, A. Plastino, O. Rosso, Phys.  Rev. Lett. {\bf 99} (2007) 154102.

\bibitem{6} E. O. Wilson, W. H. Bossert. {\it A Primer of Population Biology} (Sinauer Associates, Sunderland, MA 01375, 1971).

\bibitem{7} J.P. Gabriel, F. Saucy, L. F. Bersier, Ecological Modelling {\bf 185} (2005) 147.

\bibitem{zimm} A. D. Zimm, Comp. \& Math. Org. Theo., {\bf 11} (2005) 37.

\bibitem{toston} T. Royama, {\it Analytical Population Dynamics} (Chapman and Hall, London, 1992).

\bibitem{jaynes} E. T. Jaynes, (1957). Phys. Rev. {\bf 106}, 620 (1957); {\bf 108}, 171 (1957); IEEE Trans. Syst. Sci. \& Cyb. {\bf 4}, 227 (1968).

\bibitem{katz}  A. Katz, {\it Principles of statistical mechanics: the information theory approach} (W. H. Freeman, San Francisco, 1967).

\bibitem{prl} A. Hernando, A. Plastino, \emph{Variational Principle underlying Scale Invariant Social Systems}. Pre-print (2012).
\bibitem{epjb} A. Hernando, A.R. Plastino, A. Plastino, Eur. Phys. J. B., accepted for publication (2012).
\bibitem{ohio} Census bureau website, Government of USA, www.census.gov.
\bibitem{BM} B. H. Lavenda, {\it Nonequilibrium Statistical Thermodynamics} (John Wiley \& Sons Inc., 1985). 
\bibitem{ournets} A. Hernando, D. Villuendas, C. Vesperinas, M. Abad, A. Plastino, Eur. Phys. J. B {\bf 76}, 87 (2010).
\bibitem{clusters} C. Castellano, S. Fortunato, and V. Loreto, Rev. Mod. Phys., \textbf{81}, 591 (2009).

\bibitem{wikiBW} Wikipedia, \emph{Browser wars}. http://en.wikipedia.org/wiki/Browser\_wars.
\bibitem{w3} http://www.w3schools.com/browsers/browsers\_stats.asp
\bibitem{sc} http://gs.statcounter.com/
\bibitem{ITM} V. S. Popov, Phys. of Atom. Nuclei, {\bf 68} (2008) 686-708.
\bibitem{DFT} Parr, R. G.; Yang, W.{\it Density-Functional Theory of Atoms and Molecules} (Oxford University Press, New York. 1989)


\end{thebibliography}
\end{document}